\documentclass[twocolumn,showpacs,superscriptaddress,prb,floatfix,subequations]{revtex4}
\usepackage{amsbsy,amssymb,amsmath,bm,ulem}
\usepackage{graphicx}
\newcommand{\av}[1]{\left \langle #1 \right \rangle}

\newcommand{\cP}{\mathcal {P}}
\newcommand{\cH}{\mathcal {H}}
\newcommand{\cK}{\mathcal {K}}
\newcommand{\cL}{\mathcal {L}}
\newcommand{\cN}{\mathcal {N}}

\newcommand{\re}{\rho}
\usepackage{color}
\newcommand{\ignore}[1]{}
\newcommand{\bComment}[1]{}
\newcommand{\yComment}[1]{}
\newcommand{\dComment}[1]{}
\newcommand{\ds}[1]{}
 \renewcommand{\bComment}[1]{\textcolor{blue}{Boris: #1}}
 \renewcommand{\yComment}[1]{\textcolor{green}{Yuri: #1}}
 \renewcommand{\dComment}[1]{\textcolor{blue}{#1}}
\renewcommand{\ds}[1]{ \vspace*{0.05in} {\color{blue}\hrule
    \noindent\textsf{Daniel: #1}\hrule}\vspace*{0.05in} }
\newcommand{\eq}[1]{Eq.~(\ref{#1})}

\def\be{\begin{equation}}
\def\ee{\end{equation}}

\begin{document}
\title{ Non-Gaussian dephasing in flux qubits due to $1/f$-noise}
  \author{Y. M. Galperin}
  \email{iouri.galperine@fys.uio.no}
\affiliation{Department of
  Physics and Center of Advanced Materials and Nanotechnology,
  University of Oslo, PO Box 1048 Blindern, 0316 Oslo,
  Norway}
\affiliation{Argonne National Laboratory, 9700 S. Cass
  Av., Argonne, IL 60439, USA}
\affiliation{A. F. Ioffe
  Physico-Technical Institute of Russian Academy of Sciences, 194021
  St. Petersburg, Russia}

  \author{B. L. Altshuler}
  \affiliation{Department of Physics, Columbia University, 538
  W. 120th St., New York, NY 10027, USA}
  \affiliation{NEC Research Institute, 4 Independence Way,
          Princeton, NJ 08540, USA}
  \author{J. Bergli}
  \affiliation{Department of
  Physics, University of Oslo, PO Box 1048 Blindern, 0316 Oslo,
  Norway}
 \author{D. Shantsev}
  \affiliation{Department of
  Physics, University of Oslo, PO Box 1048 Blindern, 0316 Oslo,
  Norway}
\author{V. Vinokur}
\affiliation{Argonne National Laboratory, 9700 S. Cass
  Av., Argonne, IL 60439, USA}
%% \author{D. V. Shantsev}
%% \affiliation{Department of Physics, University of Oslo, PO Box
%% 1048 Blindern, 0316 Oslo, Norway} \affiliation{A. F. Ioffe
%% Physico-Technical Institute of Russian Academy of Sciences, 194021
%% St. Petersburg, Russia}
\date{\today}

\begin{abstract}
Recent experiments by F. Yoshihara {\em et al.} [\prl \textbf{97},
167001 (2006)]
and by K. Kakuyanagi {\em et al.} (cond-mat/0609564) provided
information on decoherence of the echo signal in
Josephson-junction flux qubits at various bias conditions. These
results were interpreted assuming a Gaussian model for the
decoherence due to $1/f$ noise. Here we revisit this problem on
the basis of the exactly solvable spin-fluctuator model
reproducing detailed properties of the $1/f$ noise interacting
with a qubit. We consider the time dependence of the echo signal
and conclude that the results based on the Gaussian assumption
need essential reconsideration.
\end{abstract}


\maketitle
%\section{Introduction}
%\paragraph{Introduction.}\label{intro}

Two remarkable works~\cite{Yoshihara2006,Kakuyanagi2006}
have appeared recently,
 reporting on 
measurements of the two-pulse echo signal in Josephson junction
flux qubits revealing its dependence of the time interval between
the pulses.  The authors interpreted their results in terms of the
recent theory \cite{Cottet2002} of the decoherence in qubits
caused by $1/f$ noise and obtained the qubit dephasing rate,
$\Gamma_\varphi$, based on the postulation of the Gaussian
statistics of the noise. 
 Although this looks like a natural starting point,
 the importance of
the  echo signal data for understanding the underlying
mechanisms of the qubits decoherence, calls for
 careful examination of  the assumptions built into
the theoretical description.

In this paper we develop a theory of the time dependence of the
echo signal making use of an exactly solvable but experimentally
realistic model.  We demonstrate that in many practical
realizations of $1/f$ noise the results based on the Gaussian
assumption need to be significantly corrected. We show that
deviation of noise statistics from the Gaussian changes
significantly {\em the time
  dependence} of the echo signal. Using  the exactly solvable
non-Gaussian {\em  spin-fluctuator model} for the $1/f$ noise we
 analyze the dependence of the echo signal on both time and the
qubit working point, and compare the obtained results with those
derived within the Gaussian approximation.

 Let us start with a brief  review the procedure used
in Refs.~\onlinecite{Yoshihara2006,Kakuyanagi2006}. 
In order to separate
the relaxation due to direct transitions between the energy levels
of the qubit, characterized by the time $T_1$, the echo
signal is expressed as $e^{-t/2T_1}\re (t)$, where 
$t$ is the
arrival time of the echo signal. 
The factor $\re (t)$  describes the pure dephasing.

If dephasing is induced by the
noise of the magnetic flux in the SQUID loop, the echo signal
 is determined by the fluctuating part of the magnetic flux through 
the loop, $\Phi (\tau)$:
\begin{equation}
  \label{eq:002}
   \re (t)=\av{\exp \left[-i  v_\Phi \left(\int_0^{t/2}
   -\int_{t/2}^{t}\right) \Phi(\tau)\,
   d\tau\right]}.
\end{equation}
 The 
  strength of the coupling between the qubit and the
noise source, $v_\Phi$, 
depends on the working point of the qubit.~\cite{Yoshihara2006,Kakuyanagi2006}

Assuming that the statistics of the fluctuations of $\Phi (t)$ is
 Gaussian  one can express
the decoherence rate through the magnetic noise spectrum,
$S_\Phi (\omega)=(1/\pi)\int_0^\infty dt\, \cos \omega t \, 
\langle\Phi(t) \Phi(0)\rangle$:
\begin{equation}
  \label{eq:003}
  \re (t)=\exp \left[-\frac{v^2_\Phi t^2}{2} \int_{-\infty}^\infty d \omega \,
  S_\Phi (\omega)\frac{\sin^4 (\omega t/4)}{(\omega t/4)^2}\right]\,.
\end{equation}
This expression is common in the theory of spin resonance and
allows one to find the decoherence rate from the noise spectrum once the
coupling, $v_\Phi$, is known. For  the  $1/f$ noise spectrum
\begin{equation}
  \label{eq:00a}
  S_\Phi (\omega)=A_\Phi/\omega\, ,
\end{equation}
one obtains
\cite{Shnirman2002,Makhlin2004,Yoshihara2006}
\begin{equation}
  \label{eq:004}
\re (t)=e^{-(\Gamma_\phi^\Phi t)^2} \, , \quad \Gamma_\phi^\Phi
\equiv
  |v_\Phi|\sqrt{A_\Phi\ln 2}\, .
\end{equation}
 The
experimental results of Ref.~\onlinecite{Kakuyanagi2006} were  fitted to
\begin{equation}
  \label{eq:005}
  \rho (t)=\exp \left[-t/2T_1-\Gamma_\phi^{0}t -
  (\Gamma_\phi^\Phi t)^2\right]\,   ,
\end{equation}
and $T_1$ and $\Gamma_\phi^{0}$ were extracted from the analysis
of the echo near the optimal point. The dephasing rate,
$\Gamma_\phi^\Phi$, was in its turn found from the fitting  and
displayed as a function of the working point.

The deficiencies of this procedure become clear when we compare
the approximate expression \eq{eq:004} with exact  equations that follow
from the spin-fluctuator model which we will now introduce.

\paragraph*{Spin-fluctuator (SF) model for $1/f$-noise. } \label{model}

One of the most common sources of the low-frequency noise is the
rearrangement of dynamic two-state defects, {\em fluctuators},
see, e.~g., book~[\onlinecite{Kogan1996}] and references therein.
Random switching of a fluctuator  between its two metastable
states (1 and 2)  produces a random telegraph noise. The process
is characterized by the switching rates $\gamma_{12}$ and
$\gamma_{21}$ for the transitions $1\to 2$ and $2\to 1$. Only the
fluctuators with the energy splitting $E$ less than
temperature, $T$, contribute to the dephasing since the
fluctuators with large level splittings are frozen in their ground
states.  As long as $E<T$ the rates $\gamma_{12}$ and $\gamma_{21}$
are of the same order of magnitude, and without loss of
generality one can assume that $\gamma_{12} \approx
\gamma_{21} \equiv \gamma/2$. The random telegraph process, $\xi
(t)$, is defined as switching between the values $\pm 1/2$ at random
times, the probability to make $n$ transitions during the time
$\tau$ being given by the Poisson distribution.
%$$P_n(\tau)=\frac{(\gamma|\tau|)^n}{n!}e^{-\gamma |\tau|} \, .$$
The correlation function of the random telegraph processes is 
$\av{\xi(t)\xi(t+\tau)}=e^{-\gamma|\tau|}/4$ and the corresponding
contribution to the noise spectrum is a Lorentzian, $\gamma/4\pi
(\omega^2 + \gamma^2)$.  Accordingly, should there
be many fluctuators coupled to the given qubit via
constants $v_i$ and having the switching rates $\gamma_i$, the
dephasing of the noise power spectrum is expressed through the sum  $\sum_i
v_i^2\gamma_i/(\omega^2+\gamma_i^2)$.
If the number of effective fluctuators is sufficiently large, the
sum over the fluctuators transforms into an integral over $v$ and
$\gamma$ weighted by the distribution function $\cP(v,\gamma)$.
Upon the assumption that the coupling constants and switching
rates are uncorrelated, the distribution density factorizes,
$\cP(v,\gamma)=\cP_v(v)\cP_\gamma(\gamma)$, and the noise spectral
function due to fluctuators reduces to
$\av{v^2}S(\omega)$, where
\begin{equation} \label{eq:007}
   \av{v^2} \equiv \int dv\, v^2\cP_v(v)\, , \quad
  S(\omega)=\frac{1}{4\pi}\int
  \frac{\gamma \cP_\gamma (\gamma)\, d\gamma }{\omega^2+\gamma^2}\, .
\end{equation}

The distribution, $\cP_\gamma (\gamma)$, is determined by the
details of the interaction between the fluctuators and their
environment, which causes their switchings.  A fluctuator viewed
as a two-level tunneling system, is characterized by two
parameters -- diagonal splitting, $\Delta$, and tunneling
coupling, $\Lambda$, the distance between the energy levels being
$E=\sqrt{\Delta^2+\Lambda^2}$ (we follow notations of
Ref.~[\onlinecite{Galperin2004}] where the model is described in
detail).  The environment is usually modeled as a boson bath,
which can represent not only the phonon field, but, e. g.,
electron-hole pairs in the conducting part of
 the system.
The external degrees of freedom are coupled with the fluctuator
via modulation of $\Delta$ and $\Lambda$, modulation of the
diagonal splitting $\Delta$ being most important. Under this
assumption the
 fluctuator-environment interaction Hamiltonian acquires the form
\begin{equation}
  \label{eq:010}
  \cH_{F-\text{env}}=\hat{c}\left(\frac{\Delta}{E}\sigma_z -
  \frac{\Lambda}{E}\sigma_x\right)\,  ,
\end{equation}
where $\hat{c}$ is an operator depending on the concrete
interaction mechanism. Accordingly, the factor $(\Lambda /E)^2$
appears in the inter-level transition rate:
\begin{equation}
  \label{eq:011}
  \gamma (E,\Lambda)=(\Lambda/E)^2\, \gamma_0 (E)\, .
\end{equation}
Here the quantity $\gamma_0$ has a meaning of the {\em maximal}
relaxation rate for fluctuators with the given energy splitting,
$E$. The coupling, $\Lambda$, depends exponentially on the
smoothly distributed tunneling action, leading to the
$\Lambda^{-1}$-like distribution of $\Lambda$, and consequently
$\cP_\gamma \propto \gamma^{-1}$.

Since only the fluctuators with $E\lesssim T$ are important
and temperatures we are interested in are low as compared to the
relevant energy scale, it is natural to assume the distribution of
$E$ to be almost constant. Denoting the corresponding density of
states in the energy space as $P_0$ we arrive at the distribution
of the relaxation rates as\cite{endnote}
\begin{equation}
  \label{eq:012a}
  \cP_\gamma
 (\gamma)=\frac{P_0T}{\gamma}\,
 \Theta(\gamma_0-\gamma)\, .
\end{equation}
The product $P_0 T$  determines the amplitude of the 1/f noise.
Indeed, the
integral $\int \cP_\gamma (\gamma)\, d \gamma$ is nothing but the
total number of thermally excited fluctuators, $\cN_T$.
Consequently,
\begin{equation}
  \label{eq:013a}
  P_0 T=\cN_T/\cL\, , \quad \cL \equiv \ln (\gamma_0/\gamma_{\min})
\end{equation}
where $\gamma_{\min}$ is the {\em minimal} relaxation rate of the
fluctuators. In the following we will assume that the number of
fluctuators is large, so that $P_0T \gg 1$ or $\cN_T \gg 2\pi \cL$.
 After that the Eq.~(\ref{eq:007}) yields:
\begin{equation}
  \label{eq:013}
  S(\omega)=\frac{ A}{\omega}\times \left\{
\begin{array}{ccc}
1&,&\omega \ll \gamma_0 \\
2\gamma_0/\pi\omega &,& \omega \gg \gamma_0 \end{array}\, , \right.
   \quad A\equiv \frac{1}{8}P_0T\, .
\end{equation}
We see that the SF model reproduces the $1/f$ noise power spectrum
(\ref{eq:00a}) for $\omega\ll\gamma_0$.
The crossover from $\omega^{-1}$ to $\omega^{-2}$ behavior at
$\omega \sim \gamma_0$ follows from the existence of a 
 maximal switching rate $\gamma_0$. Below we will see that this 
crossover modifies 
the time dependence of the echo signal at times $t\ll\gamma_0^{-1}$.

The SF model has previously
been used for description of effects of noise in various
systems~\cite{Ludviksson1984,Kogan1984,Kozub1984,Galperin1991,Galperin1994,Hessling1995}
and was recently applied to analysis of decoherence in charge
qubits.~\cite{Paladino2001,Paladino2003,Galperin2004,Falci2003,Falci2004,Falci2005,Galperin2006,Martin2006,Bergli2006}
Quantum aspects of the model were addressed in
Ref.~\onlinecite{DiVincenzo2005}.  These studies, demonstrated, in
particular, that the SF model is suitable for the study of
non-Gaussian effects and that these may be essential in certain situations.

Now we are ready to  analyze consequences of the upper cutoff and
the effect of non-Gaussian noise,
and through this identify the validity region for the prediction of
Eq.~(\ref{eq:004}). 

In the following we will express the fluctuation
of the magnetic flux as a sum of the contributions of the statistically
independent fluctuators, $\Phi (t)=\sum_i b_i\xi_i (t)$, where $b_i$
are partial amplitudes while $\xi_i(t)$ are random telegraph
processes. Consequently, we express the product $v_\Phi^2 A_\Phi$ as
$\bar{v}^2A$ where $\bar{v} \equiv \sqrt{\av{v_\Phi^2b^2}}$. In
other words, we include the amplitude of the magnetic noise in the
effective coupling constant.

\paragraph{Echo signal in the Gaussian approximation.} \label{gaussian}

Substitution of the distribution (\ref{eq:012a}) into
Eq.~(\ref{eq:003}) yields for $ \cK(t) \equiv -\ln \re (t)$
\begin{equation}
  \label{eq:014}
  \cK_g(t) =\bar{v}^2 At^2 \times \left\{ \begin{array}{ccl}
\gamma_0 t/6 &,& \gamma_0 t \ll 1 \\
\ln 2 &,& \gamma_0 t \gg 1 \, . \end{array} \right.
\end{equation}
%Here $\bar{v}\equiv \av{v^2b^2}^{1/2}$. 
The subscript $g$ means that
this result is obtained in the Gaussian approximation. One sees
immediately that the crossover between the $\omega^{-1}$ and
$\omega^{-2}$ behaviors in the noise spectrum does not
affect the echo signal at long times  $t
\gg \gamma_0^{-1}$. However, at small times the decay
decrement acquires an extra factor $\gamma_0 t$, which is nothing
but the probability for a typical fluctuator to change its state
during the time $t$. As a result, even in the the Gaussian
approximation at small times $\cK_g(t) \propto t^3$ (replacing
$\cK_g(t) \propto t^2$ behavior).

\paragraph{Non-Gaussian theory:} \label{exact}

The echo signal given by \eq{eq:002} can be calculated exactly
using the method of stochastic differential equations, if the
fluctuating quantity $\Phi$ is a single random telegraph process,
see Ref.~\onlinecite{Galperin2004} and references therein. The
method was developed in the contexts of spin
resonance,~\cite{Klauder1962} spectral diffusion in
glasses,~\cite{Black1977,Hu1977,Maynard1980,Laikhtman1985} and
single molecular spectroscopy in disordered
media.\cite{Moerner1994,Geva1996,Moerner1999,Barkai2001} Averaging
in \eq{eq:002} is performed over random realizations of $\Phi$ and
its initial states and reflects the conventional experimental
procedure\cite{Yoshihara2006,Kakuyanagi2006}  where the observable
signal is accumulated over numerous repetitions of the same
sequence of inputs.

For a single fluctuator with switching rate $\gamma/2$ and coupling
$v$ we find \cite{Galperin2006}
% \begin{eqnarray}
\[ \rho_1(t)=\frac{e^{-\gamma t/2}}{2\mu^2}\left [ (1+\mu)
 e^{\mu\gamma t/2}+(1-\mu) e^{-\mu\gamma t/2} 
% &&\left. -
 - \frac{2v^2}{\gamma^2}  \right]\, , % \quad \mu=\sqrt{1-(v/\gamma)^2}\,
 \label{eq:04}
\]
%\end{eqnarray}
where $\mu=\sqrt{1-(v/\gamma)^2}$. 
In the appropriate limits this can be expanded to give
 \begin{equation}
   \label{eq:017}
   -\ln\rho_1(t|v,\gamma)\approx \left\{\begin{array}{ccl}
v^2 \gamma t^3/48 &,& t^{-1} \gg \gamma, v  \\
\gamma t/2 &,& \gamma,t^{-1} \ll v\\
v^2 t /2\gamma &,&  \gamma \gg v, t^{-1}\, .
\end{array} \right.
 \end{equation}
Note that the
last limiting case here is similar to the motional narrowing of
spectral lines well know in physics of spin
resonance.~\cite{Klauder1962}

In the case of many fluctuators producing $1/f$-like noise, $\cN_T
\gg 1$,  the sum of the contributions from individual fluctuators
 gives the echo
decay decrement
\begin{equation}
  \label{eq:016}
  \cK_{\text{sf}}(t)= -\int dv\, \cP_v(v)  \int d\gamma \, \cP_\gamma
  (\gamma)\,\ln\rho_1(t|v,\gamma) \, .
\end{equation}
Let us assume that the distribution of $v$ is a sharp function
centered at some value $\bar{v}$.  This reduces integration over
$v$ to mere replacing $v \to \bar{v}$ in the expressions for
$\re(v,t|\gamma)$. This approximation is valid as long as
$\av{v^2}$ is finite. This is seemingly the case, e. g.,
for magnetic noise induced by tunneling of
vortices between different pinning centers within the SQUID
loop. The case of divergent $\av{v^2}$ is considered in
Refs.~\onlinecite{Galperin2004,Galperin2006,Schriefl2006}. 
Using then  (\ref{eq:012a}) for the distribution function $\cP_\gamma$
and using the appropriate terms from Eq. (\ref{eq:017}) in 
Eq. (\ref{eq:016}) we find the time dependence of the 
logarithm of the echo signal, $K_{\text{sf}}(t)$. For $\bar{v}
  \ll \gamma_0$
 \begin{equation}
   \label{eq:018}
   \cK_{\text{sf}}(t) \approx \left\{\begin{array}{ccll}
\gamma_0A\bar{v}^2  t^3/6 &,& &t \ll \gamma_0^{-1},  \\
\ln 2 \bar{v}^2 A t^2 &,&   \gamma_0^{-1}\ll &t\ll\bar{v}^{-1},\\
%\gamma_0 At}&,&  v \gg \gamma_0 \gg 1 \\
\alpha\bar{v}A t &,&   \bar{v}^{-1}\ll &t.
\end{array} \right.
 \end{equation}
where $\alpha \approx 6$.

At small times $t\ll\bar{v}$ we arrive at the same result as in the
{\em  properly
treated} Gaussian approach, Eq (\ref{eq:014}).
 However, at large times, $t\gg\bar{v}^{-1}$, the exact result
 {\em dramatically differs} from the prediction of the Gaussian
 approximation. To understand the reason,  notice that
for $\cP_\gamma (\gamma) \propto 1/\gamma$, the decoherence is dominated by
fluctuators  with $\gamma \approx v$. The physical
reason for that is clear: very ``slow'' fluctuators produce slow
varying fields, which are effectively refocused in course of the
echo experiment, while the influence of too ``fast'' fluctuators
is reduced due to the effect of motional narrowing. As shown in
Ref.~\onlinecite{Galperin2006}, only the fluctuators with $v \ll
\gamma$ produce Gaussian noise. Consequently, the noise in this
case is essentially {\em non-Gaussian}. Only at short 
 times $t\ll\bar{v}^{-1}$ when these
most important fluctuators did not yet have time to switch, and 
only the faster fluctuators contribute, is  the Gaussian approximation 
 valid.

For $\bar{v} \gg \gamma_0$ we find 
\begin{equation}
   \label{eq:018a}
   \cK_{\text{sf}}(t) \approx \left\{\begin{array}{ccl}
\gamma_0 A\bar{v}^2 t^3/6 &,& t \ll  \bar{v}^{-1},  \\
4\gamma_0 A t &,&    t\gg \bar{v}^{-1}.
\end{array} \right.
 \end{equation}
In this case all fluctuators have $v\gg\gamma$, hence the result 
at long times again differs significantly from the Gaussian result 
Eq. (\ref{eq:014}).

\paragraph*{Discussion:} \label{discussion}

\begin{figure}
  \begin{center}
\includegraphics[width=8cm]{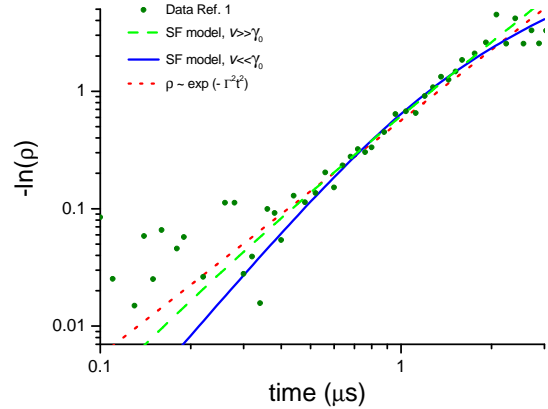}
  \end{center}
\caption{Dephasing component of the echo measurements replotted from fig 4a of 
 Ref. \protect\onlinecite{Yoshihara2006} away from the optimal point. The curves show 
fits to the SF model, eqs. 
(\ref{eq:018}) and  (\ref{eq:018a}) and to the $\rho=e^{-\Gamma^2 t^2}$
law.   
The fitting took into account all data points including those 
that fall outside the range of the plot (e. g. those with $\rho>1$).
\label{f1}}
\end{figure}

Here we apply the results obtained above to analyze quantitatively 
the decoherence of the flux qubit.
Fig. \ref{f1} shows  the time dependence of the echo signal measured in
 Ref. \onlinecite{Yoshihara2006}
and a fit based on the SF model. We consider first the case of 
$\bar{v} \gg \gamma_0$ where the SF model predicts a crossover from 
$t^3$ to $t$ dependence, Eq. (\ref{eq:018a}). We replace the exact result by 
the interpolation formula, 
$\cK_{\text{sf}}(t) =\gamma_0 At^3/(6 \bar{v}^{-2}+t^2/4)$.
Figure \ref{f1} also shows the commonly used fit $\rho=e^{-\Gamma^2 t^2}$,
 which is represented by a straight
line with slope 2. The fit to the SF model seems slightly better, or at 
least equally good.
From this fit we can extract the average change in the qubit energy splitting 
$E_{01}$
due to a flip of one fluctuator, $\bar{v}\approx 4 \mu$s$^{-1}$.
It allows us to evaluate the change of flux in the qubit loop induced by 
a fluctuator flip, $b=\bar{v}/v_\Phi$, since 
$v_\Phi=(1/\hbar)\partial E_{01}/\partial\Phi$ was measured in Ref. 
\onlinecite{Yoshihara2006}. Using the experimental values for all 
parameters we get $b\approx 4.2\times10^{-6}\Phi_0$, where $\Phi_0$ is the 
flux quantum while the deviation from the optimal working point 
was $\approx10^{-3}\Phi_0$. Hence the flip of one fluctuator 
changes the qubit energy splitting by 0.4\%.
The fit also determines the value of the product $\gamma_0A$ and within 
our assumption $\gamma_0<v$ we find a lower estimate for 
the flux noise amplitude $\sqrt{A_\Phi}/\Phi_0>1.3\times10^{-6}$.

Similarly, if $\bar{v} \ll \gamma_0$ we can fit to Eq. 
(\ref{eq:018}), using the interpolation formula 
$\cK_{\text{sf}}(t) =\bar{v}^2 At^3/(6\gamma_0^{-1}+t/\ln2)$. 
We do not include here the $\cK_{\text{sf}}\propto t$ behavior at large 
$t$ since it corresponds to $\cK_{\text{sf}}>A\gg1$. 
The fitting gives $\gamma_0\approx 30$~$\mu$s$^{-1}$, which is then also an 
upper estimate for $\bar{v}$. We also get 
$\sqrt{A_\Phi}/\Phi_0=1.05\times10^{-6}$. 

The quality of the data does not allow us to determine which of the 
cases $\bar{v} \gg \gamma_0$ or $\bar{v} \ll \gamma_0$ is realized in 
the experiments. However, in both cases the value of $A_\Phi$ is similar
and close to the value obtained in the Gaussian approach.~\cite{Yoshihara2006}
Besides, even without determining which case is realized we obtain that 
both $\bar{v}$ and $\gamma_0$ can not be larger than $30$~$\mu$s$^{-1}$.
This gives an upper estimate for the change of flux in 
the qubit loop as one fluctuator flips, $b < 3\times 10^{-5}\Phi_0$.
Since $\gamma_0$ is expected to grow with temperature, we believe that 
it would be instructive to performe similar analysis of the echo 
decay at different temperatures.

\paragraph*{Conclusions:}
By introducing the spin fluctuator model for $1/f$ noise in the 
qubit level splitting we have 
determined the time dependence of the echo signal. We show that 
the standard quadratic time dependence in the Gaussian approximation 
Eq.~(\ref{eq:004}) has a limited range of applicability, 
and 
$t^3$ or $t$
dependencies are found beyond this range. 
Fitting to the SF model also allows us to determine the 
strength of {\it individual} fluctuators, and for the flux qubits 
reported in Ref. \onlinecite{Yoshihara2006} the change of 
flux in the qubit loop due to the flip of one fluctuator 
was found to be $b < 3\times 10^{-5}\Phi_0$.

\acknowledgments This work was partly supported by the Norwegian
Research Council, Funmat@UiO, and by the U. S. Department of  Energy Office of
Science through contract No. DE-AC02-06CH11357.
We are thankful to Y. Nakamura for helpful discussions and for providing experimental data.

\end{document}